# Structural Analysis and Superconducting Properties of F-Substituted NdOBiS$_2$ Single Crystals

Masanori Nagao[1,2*], Satoshi Demura[2], Keita Deguchi[2], Akira Miura[1], Satoshi Watauchi[1],

Takahiro Takei[1], Yoshihiko Takano[2], Nobuhiro Kumada[1], and Isao Tanaka[1]

[1]*University of Yamanashi, 7-32 Miyamae, Kofu 400-8511, Japan*

[2]*National Institute for Materials Science, 1-2-1 Sengen, Tsukuba 305-0047, Japan*

(Received)

## Abstract

F-substituted NdOBiS$_2$ superconducting single crystals were grown using CsCl/KCl

flux. This is the first example of the single-crystal growth of a BiS$_2$-based

superconductor. The obtained single crystals had a plate-like shape with a size of 1-2

mm and a well-developed *ab*-plane. The crystal structure of the grown crystals was

determined by single-crystal X-ray diffraction analysis to be the tetragonal space group

*P4/nmm* (#129) with $a = 3.996(3)$ Å and $c = 13.464(6)$ Å. The chemical formula of the grown crystals was approximately Nd$_{0.98\pm0.06}$O$_{0.7\pm0.1}$F$_{0.3\pm0.1}$Bi$_{0.98\pm0.04}$S$_2$, and Cs, K, and Cl were not detected in the grown crystals by electron probe microanalysis. The grown crystals had a critical temperature of approximately 5 K. The superconducting anisotropy of the single crystals was estimated to be about 30 from the effective mass model and the upper critical field.

*E-mail address: mnagao@yamanashi.ac.jp

*Corresponding Author

Masanori Nagao

Postal address: University of Yamanashi, Center for Crystal Science and Technology

Miyamae 7-32, Kofu 400-8511, Japan

Telephone number: (+81)55-220-8610

Fax number: (+81)55-254-3035

**Main text**

Since the discovery of the $BiS_2$-based superconductors $Bi_4O_4S_3$,[1] $RO_{1-x}F_xBiS_2$ (R=La, Ce, Pr, Nd, Yb),[2-6] and $Sr_{1-y}La_yFBiS_2$,[7] much effort has been focused on the growth of their single crystals. Single crystals of $BiS_2$-based superconductors are attracting much attention because they will enable the analysis of their intrinsic physical properties. However, bulk-size single crystals of $BiS_2$-based superconductors have never been grown. Alkali chloride fluxes are useful for obtaining single crystals. So far, single crystals of $Bi_2Sr_2CaCu_2O_x$ and $Bi_2Sr_2Ca_2Cu_3O_y$ high-$T_c$ superconductors have been grown using KCl flux.[8-10] However, $BiS_2$-based superconductors cannot be synthesized using KCl flux in a vacuum quartz tube because KCl would corrode the quartz tube at high temperatures. Therefore, a mixture of CsCl and KCl (CsCl/KCl) is a promising flux as it exhibits little reactivity to quartz tubes at high temperatures.[11-13]

In this paper, we focus on the compound $NdO_{1-x}F_xBiS_2$ (x=0.1–0.7), which was found to have a maximum $T_c$ of 5.6 K at x=0.3.[5] We report the successful growth of F-substituted $NdOBiS_2$ single crystals using CsCl/KCl as the flux with a eutectic temperature of approximately 616 °C in a quartz tube sealed in vacuum. For the first time, the crystal structure of the grown crystals was determined by single-crystal X-ray diffraction (XRD) analysis. The composition and superconducting properties were also investigated to characterize the properties of the single crystals.

Single crystals of F-substituted $NdOBiS_2$ were grown in a quartz tube sealed in vacuum by a high-temperature flux method. The raw materials used were $Nd_2S_3$, Bi, $Bi_2S_3$, $Bi_2O_3$, $BiF_3$, CsCl, and KCl.

The raw materials were weighed with a nominal composition of $NdO_{0.7}F_{0.3}BiS_2$. The mixture of the raw materials (0.8 g) and CsCl/KCl flux (3.0 g) were mixed using a mortar, and sealed in a quartz tube in vacuum. The molar ratio of the CsCl/KCl flux was CsCl:KCl=5:3. The mixed powder was then heated at 800 °C for 10 h followed by slow cooling to 600 °C at a rate of 0.5 °C/h. The sample was then furnace-cooled from 600 °C. After cooling to room temperature, the quartz tube was opened in air atmosphere, and distilled water was added to the quartz tube. The product was then filtered and washed with distilled water.

The crystal structure and composition of the single crystals were evaluated by XRD analysis using CuK$\alpha$ radiation, scanning electron microscopy (SEM), and electron probe microanalysis (EPMA). Single-crystal XRD data were collected using a Rigaku XTALAB-MINI diffractometer with graphite monochromated MoK$\alpha$ radiation. The data were corrected for the Lorentz and polarization effects. The crystal structure was solved and refined using computer programs from the Crystal Structure crystallographic software package.[14] Details of the data collection and refinement are summarized in Table I. The crystal structure was drawn by VESTA computer software.[15] The temperature dependence of magnetization was measured by a superconducting quantum interface device (SQUID) magnetometer with an applied field of 10 Oe. The transport properties of the single crystals were measured by the standard four-probe method with constant current density ($J$) using a Quantum Design Physical Property Measurement System (PPMS). The transition temperature corresponding to the onset of superconductivity ($T_c^{onset}$) was defined as the point at which deviation from linear behavior is observed in the normal conducting state in the resistivity-temperature ($\rho$-$T$) characteristics. The zero resistivity ($T_c^{zero}$) was determined considering the criterion of a resistivity of 0.1 $\mu\Omega$cm in the $\rho$-$T$ characteristics. We measured the angular ($\theta$) dependence of resistivity ($\rho$) in the flux liquid state under various magnetic fields ($H$) and calculated the superconducting anisotropy ($\gamma_s$) using the effective mass model.[16-18]

Figure 1 shows a typical SEM image of a F-substituted $NdOBiS_2$ single crystal. The obtained single crystals had a plate-like shape of side 1.0-2.0 mm and 10-20 $\mu$m thickness. We did not measure the in-plane orientation of the grown crystals. Figure 2 shows an XRD pattern of a well-developed plane of a F-substituted $NdOBiS_2$ single crystal. The grown single crystals were identified to have the $NdOBiS_2$ structure from the XRD pattern. The presence of only 00$l$ diffraction peaks of the $NdOBiS_2$ structure indicates that the $ab$-plane is well-developed.

The structural analysis showed that the crystal structure was the tetragonal space group $P4/nmm$ (#129) with $a = 3.996(3)$ Å, and $c = 13.464(6)$ Å. Structural refinement was performed on $NdOBiS_2$ because it is difficult to distinguish between O and F. In the initial refinement, the occupations of Nd and Bi were refined and found to be unity within the error. Thus, no defects of Nd and Bi sites were detected. The final refinement was performed with fixed occupancies of unity for Nd, Bi, S, and O. The obtained atomic coordinates are summarized in Table II. As shown in Fig. 3, the crystal structure of $NdOBiS_2$ is composed of stacked $Nd_2O_2$ layers and $Bi_2S_4$ layers, which is almost the same as the $LaOBiS_2$ structure.[2] The composition of the F-substituted $NdOBiS_2$ single crystals was estimated to be approximately $Nd_{0.98\pm0.06}O_{0.7\pm0.1}F_{0.3\pm0.1}Bi_{0.98\pm0.04}S_2$ by EPMA. The obtained values were normalized using S = 2. Nd and Bi were measured to a precision of two decimal places. On the other hand, O and F are accurate to one decimal place. These values are consistent with the crystal structure analysis within the analytical error. The measured values of the lattice constants $a$ and $c$ in the grown crystals are comparable to those of a polycrystalline powder sample with a F content of $0.3\pm0.1$.[5)] Cs, K, and Cl were not detected in the grown single crystals by quantitative analysis using EPMA with a minimum sensitivity limit of 0.1 wt%.

Figure 4 shows the temperature dependence of magnetic susceptibility for a $Nd_{0.98\pm0.06}O_{0.7\pm0.1}F_{0.3\pm0.1}Bi_{0.98\pm0.04}S_2$ single crystal. The Meissner effect was confirmed from the magnetic susceptibility between 2 and 20 K. Figure 5 shows the $\rho$-$T$ characteristics along the $ab$-plane of a $Nd_{0.98\pm0.06}O_{0.7\pm0.1}F_{0.3\pm0.1}Bi_{0.98\pm0.04}S_2$ single crystal. A superconducting transition was observed below 6 K. $T_c^{onset}$ is 5.5 K and $T_c^{zero}$ was observed at 5.0 K. Resistivity slightly decreases with decreasing temperature down to 6 K, which is metallic behavior. This is different from the property of polycrystalline samples of semiconducting behavior down to approximately 6 K.[5,6,19)] We assume that intrinsic property of $Nd_{0.98\pm0.06}O_{0.7\pm0.1}F_{0.3\pm0.1}Bi_{0.98\pm0.04}S_2$ single crystal is metallic, and that the transport property of the polycrystalline sample is semiconducting owing to the existence of grain boundaries. Figure 6 shows the temperature dependence of resistivity below 10 K under a magnetic field ($H$) of 0-9.0 T parallel to the (a) $ab$-plane and (b) $c$-axis. The resistivity in the normal state exhibits metallic behavior under a magnetic field parallel to the $ab$-plane. On the other hand, an upturn is observed in the resistivity under a magnetic field of more than 0.5 T parallel to the $c$-axis, which appears to be semiconducting behavior. Further investigation is underway to clarify this phenomenon.

The suppression of the critical temperature under the magnetic field applied parallel to the $c$-axis is more significant than that for a field parallel to the $ab$-plane. The field dependences of $T_c^{onset}$ and $T_c^{zero}$ under the magnetic fields ($H$) parallel to the $ab$-plane ($H//ab$-plane) and $c$-axis ($H//c$-axis) are plotted in Fig. 7. The linear extrapolations of $T_c^{onset}$ for the case of $H//ab$-plane and $H//c$-axis approach values of 42 and 1.3 T, respectively. The upper critical fields $H^{//ab}_{C2}$ and $H^{//c}_{C2}$ are predicted to be less than 42 and 1.3 T, respectively. Using linear fitting to the $T_c^{zero}$ data, the irreversibility fields $H^{//ab}_{irr}$ and $H^{//c}_{irr}$ are found to be 16 and 0.64 T, respectively. This result indicates that the F-substituted NdOBiS$_2$ superconductor has high anisotropy. We previously estimated the superconducting anisotropy of Nd$_{0.98\pm0.06}$O$_{0.7\pm0.1}$F$_{0.3\pm0.1}$Bi$_{0.98\pm0.04}$S$_2$ using a single crystal. The angular ($\theta$) dependence of resistivity ($\rho$) was measured at various magnetic fields ($H$) in the flux liquid state to estimate the superconducting anisotropy ($\gamma_s$) of a grown Nd$_{0.98\pm0.06}$O$_{0.7\pm0.1}$F$_{0.3\pm0.1}$Bi$_{0.98\pm0.04}$S$_2$ single crystal, as reported in Refs. 16 and 17. The reduced field ($H_{red}$) is evaluated using the following equation for the effective mass model:

$$H_{red} = H(\sin^2\theta + \gamma_s^{-2}\cos^2\theta)^{1/2}, \qquad (1)$$

where $\theta$ is the angle between the $ab$-plane and the magnetic field.[18] $H_{red}$ is calculated from $H$ and $\theta$. The superconducting anisotropy ($\gamma_s$) was estimated from the best scaling for the $\rho$-$H_{red}$ relations. Figure 8 shows the angular ($\theta$) dependence of resistivity ($\rho$) at various magnetic fields ($H$ = 0.1-9.0 T) in the flux liquid state for a Nd$_{0.98\pm0.06}$O$_{0.7\pm0.1}$F$_{0.3\pm0.1}$Bi$_{0.98\pm0.04}$S$_2$ single crystal. Small dips were observed in the $\rho$-$\theta$ curves around the $H$//$c$-axis at magnetic fields of less than 0.6 T. These dips presumably originated from small cracks, which behave as weak pinning sites in the single crystal. The $\rho$-$\theta$ curve had twofold symmetry. The $\rho$-$H_{red}$ scaling obtained from the $\rho$-$\theta$ curves in Fig. 8 using Eq. (1) is shown in Fig. 9. The scaling is obtained by taking $\gamma_s = 30$, shown in Fig. 9(a). However, this plot deviates from the scaling at a higher field of more than 2.5 T as shown in Fig. 8. As described before, $H^{//c}_{C2}$ was predicted to be less than 1.3 T from Fig. 7. Assuming that $\gamma_s$ diverges at magnetic field of more than 1.3 T, we perform $\rho$-$H_{red}$ scaling only using data from Fig. 8 obtained at magnetic fields of less than or equal to 1.0 T, as shown in Fig. 9(b). The best scaling for this set of data was obtained by taking $\gamma_s = 9$. However, when $\gamma_s$ was more than 9, the $\rho$-$H_{red}$ scaling was almost saturated at magnetic fields of less than or equal to 1.0 T. On the other hand, the superconducting anisotropy evaluated from the ratio of the upper critical field using the equation:

$$\gamma_s = H^{//ab}_{C2}/H^{//c}_{C2} = \xi_{ab}/\xi_c \quad (\xi\text{: coherence length}) \tag{2}$$

was 32.3. We assume that the superconducting anisotropy was estimated to be approximately 30.

We have successfully grown F-substituted NdOBiS$_2$ single crystals using CsCl/KCl flux. Plate-like F-substituted NdOBiS$_2$ single crystals with a size of about 1.0-2.0 mm were obtained. The structure of the grown crystals was determined to be the tetragonal space group *P4/nmm* (#129) with *a* = 3.996(3) Å and *c* = 13.464(6) Å. The composition of the F-substituted NdOBiS$_2$ single crystal was Nd$_{0.98\pm0.06}$O$_{0.7\pm0.1}$F$_{0.3\pm0.1}$Bi$_{0.98\pm0.04}$S$_2$. $T_c^{onset}$ and $T_c^{zero}$ for the single crystal were 5.5 and 5.0 K, respectively. The values of $\gamma_s$ were determined to be 30 and 32.3 using the effective mass model and the upper critical field [Eq. (2)], respectively. The CsCl/KCl flux method is a powerful way to obtain chalcogenide single crystals.

**Acknowledgment**

The authors would like to thank Drs. M. Fujioka and H. Okazaki of the National Institute for Materials Science for useful discussions.

**Figure captions**

FIG. 1. SEM image of F-substituted $NdOBiS_2$ single crystal.

FIG. 2. XRD pattern of well-developed plane of F-substituted $NdOBiS_2$ single crystal.

FIG. 3. (Color online) Crystal structure of $NdOBiS_2$. The dashed line indicates the unit cell.

FIG. 4 Temperature dependence of magnetic susceptibility for $Nd_{0.98\pm0.06}O_{0.7\pm0.1}F_{0.3\pm0.1}Bi_{0.98\pm0.04}S_2$ single crystal under zero-field cooling (ZFC) and field cooling (FC).

FIG. 5. Resistivity–temperature ($\rho$-$T$) characteristics along the $ab$-plane of $Nd_{0.98\pm0.06}O_{0.7\pm0.1}F_{0.3\pm0.1}Bi_{0.98\pm0.04}S_2$ single crystal. The inset is an enlargement of the superconducting transition.

FIG. 6. (Color online) Temperature dependence of resistivity for $Nd_{0.98\pm0.06}O_{0.7\pm0.1}F_{0.3\pm0.1}Bi_{0.98\pm0.04}S_2$ under magnetic fields of 0-9.0 T (0, 0.1, 0.2, 0.3, 0.4, 0.5, 0.6, 0.7, 0.8, 0.9, 1.0, 1.5, 2.0, 2.5, 3.0, 5.0, 7.0, and 9.0 T) parallel to the (a) $ab$-plane and (b) $c$-axis.

FIG. 7. Field dependences of $T_c^{onset}$ and $T_c^{zero}$ under magnetic fields ($H$) parallel to the $ab$-plane ($H$//$ab$-plane) and $c$-axis ($H$//$c$-axis). The lines are linear fits to the data. The inset is an enlargement of the lower-field region.

FIG. 8. (Color) Angular $\theta$ dependence of resistivity $\rho$ in flux liquid state at various magnetic fields (bottom to top, 0.1, 0.2, 0.3, 0.4, 0.5, 0.6, 0.7, 0.8, 0.9, 1.0, 1.5, 2.0, 2.5, 3.0, 5.0, 7.0, and 9.0 T) for $Nd_{0.98\pm0.06}O_{0.7\pm0.1}F_{0.3\pm0.1}Bi_{0.98\pm0.04}S_2$ single crystal.

FIG. 9. (Color) (a) Data in Fig. 8 after scaling of angular $\theta$ dependence of resistivity $\rho$ at a reduced magnetic field of $H_{red} = H(\sin^2\theta + \gamma_s^{-2}\cos^2\theta)^{1/2}$. (b) Best fit for the data for $H = 0.1$-1.0 T.

TABLE I. Crystal data and intensity collection of F-substituted $NdOBiS_2$ single crystal.

| | |
|---|---|
| Crystal system | Tetragonal |
| Space group | *P4/nmm* (No. 129), Z = 2 |
| Lattice parameters (Å) | a = 3.996(3) |
| | c = 13.464(6) |
| Volume (Å$^3$) | 215.0(4) |
| Diffractometer | XTALAB-MINI |
| Radiation | Graphite monochromated MoK$\alpha$ |
| | ($\lambda$ = 0.71075 Å) |
| Temperature (°C) | Room temperature |
| $\mu$ (MoK$\alpha$) (cm$^{-1}$) | 534.884 |
| No. of reflections measured | Total: 2128 |
| | Unique: 191 (R$_{int}$ = 0.141) |
| Corrections refinement method | Lorentz polarization |
| No. observations (all reflections) 191 | Full-matrix least-squares on F$^2$ |
| No. variables | 16 |
| Reflection/parameter ratio | 11.94 |
| Residuals: R1 (I>2.00σ(I)) | 0.0458 |
| Residuals: R (all reflections) | 0.0471 |
| Residuals: wR2 (all reflections) | 0.1276 |
| Goodness-of-fit indicator | 0.940 |

TABLE II. Atomic coordinates for F-substituted NdOBiS$_2$. All occupancies were fixed to 1.

| Site | $x$ | $y$ | $z$ | *$B_{eq}$ (Å$^2$) |
|------|-----|-----|-----|-------------------|
| Bi1 | 1/4 | 1/4 | 0.87407(9) | 1.253 |
| Nd1 | -1/4 | -1/4 | 0.59470(14) | 1.05(5) |
| S1 | -1/4 | -1/4 | 0.8800(8) | 1.69(14) |
| S2 | 1/4 | 1/4 | 0.6860(6) | 1.06(12) |
| O/F | -3/4 | -1/4 | 1/2 | 0.8(3) |

*$B_{eq}$: Equivalent isotropic atomic displacement parameter

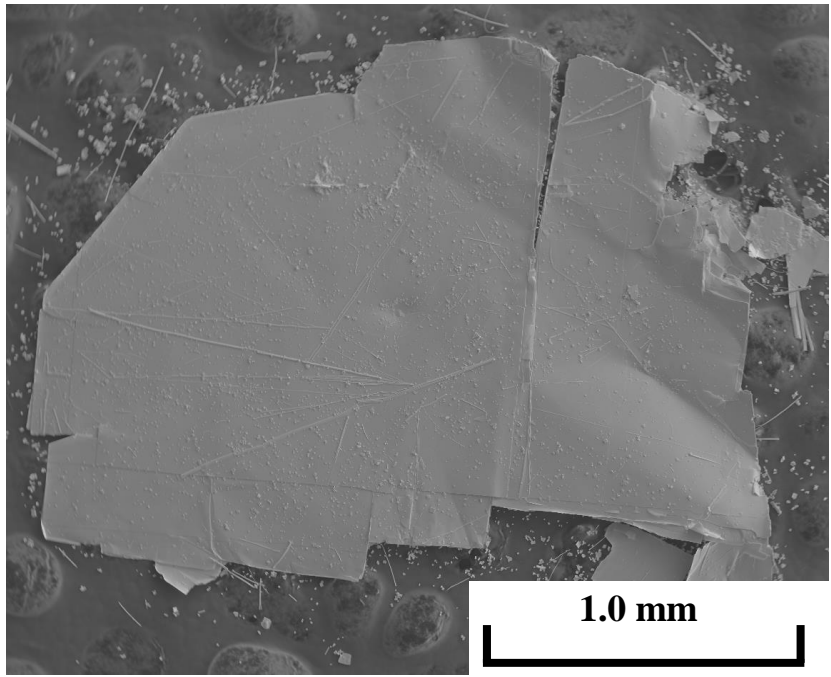

1.0 mm

**Figure 1**

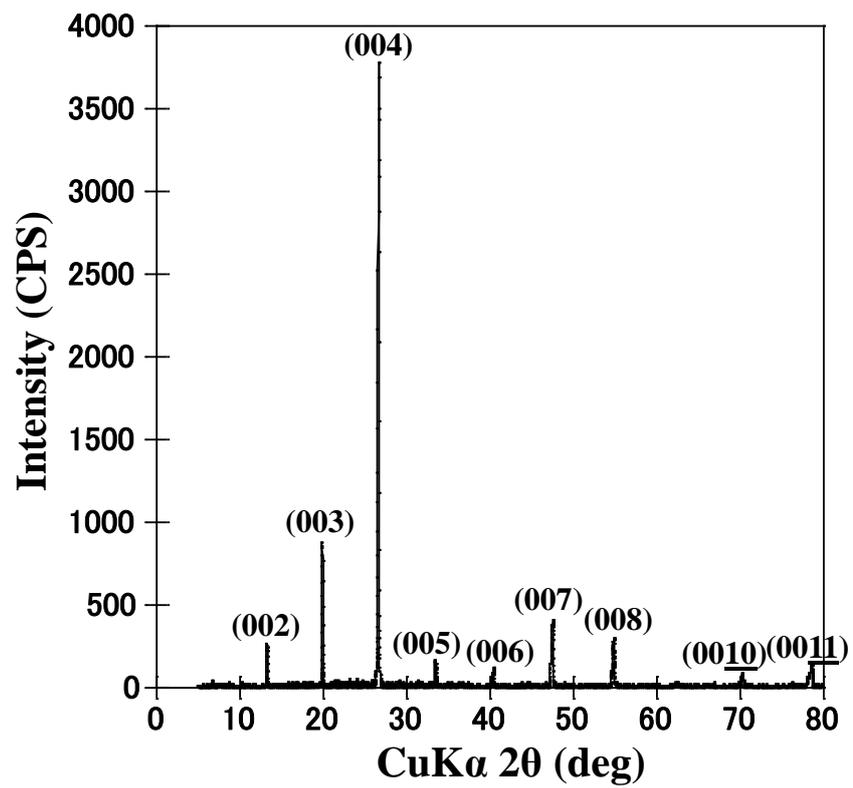

**Figure 2**

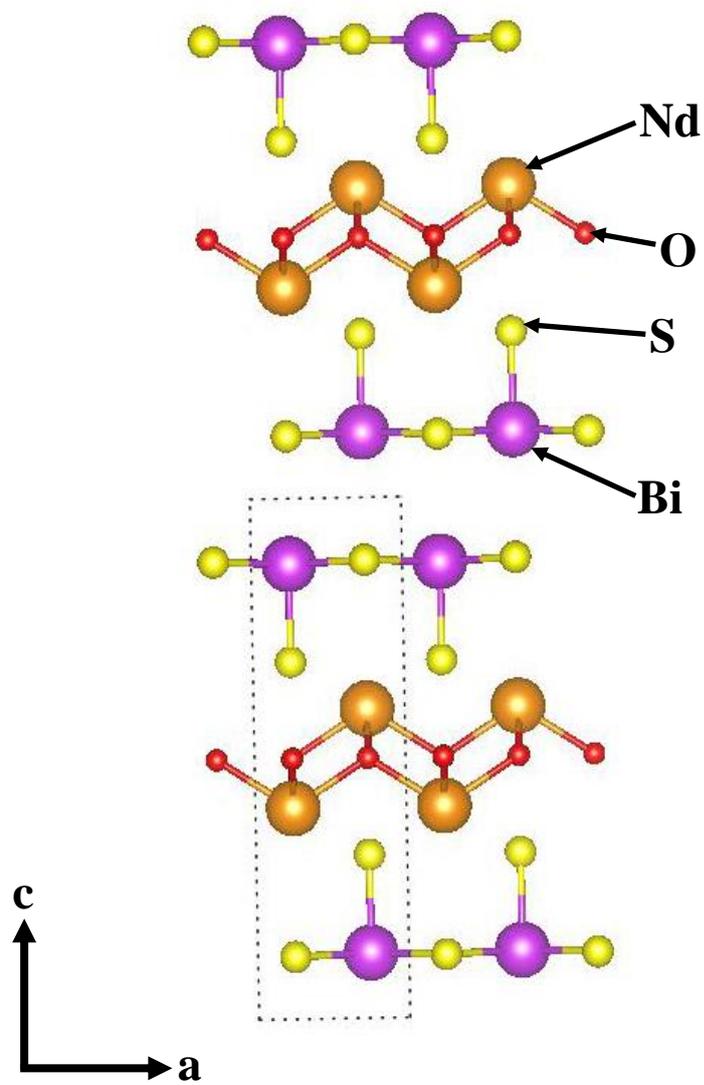

**Figure 3 (Color online)**

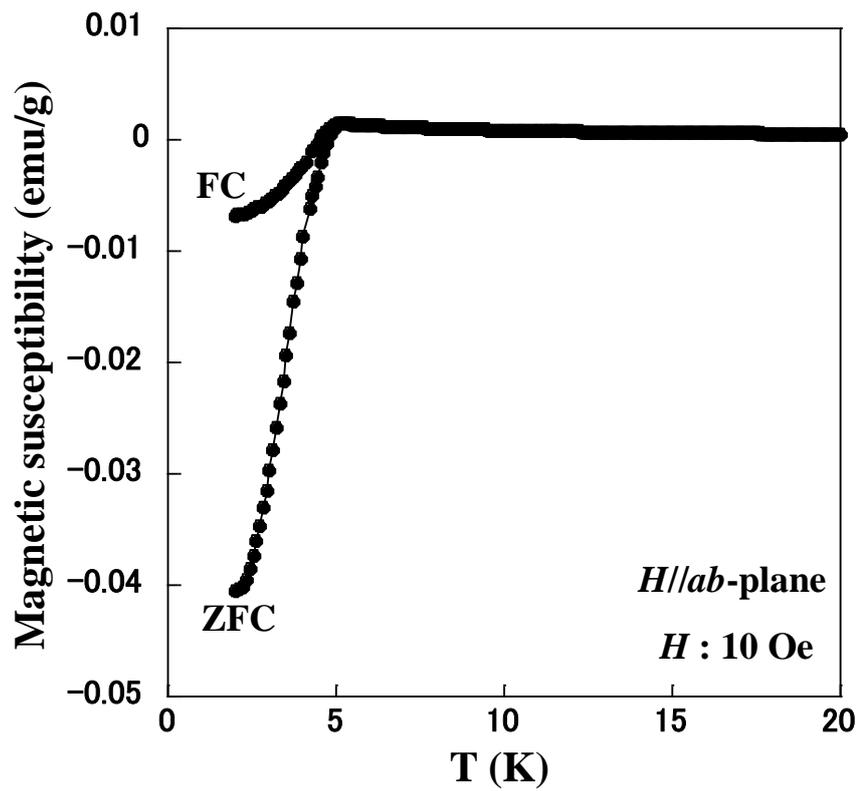

**Figure 4**

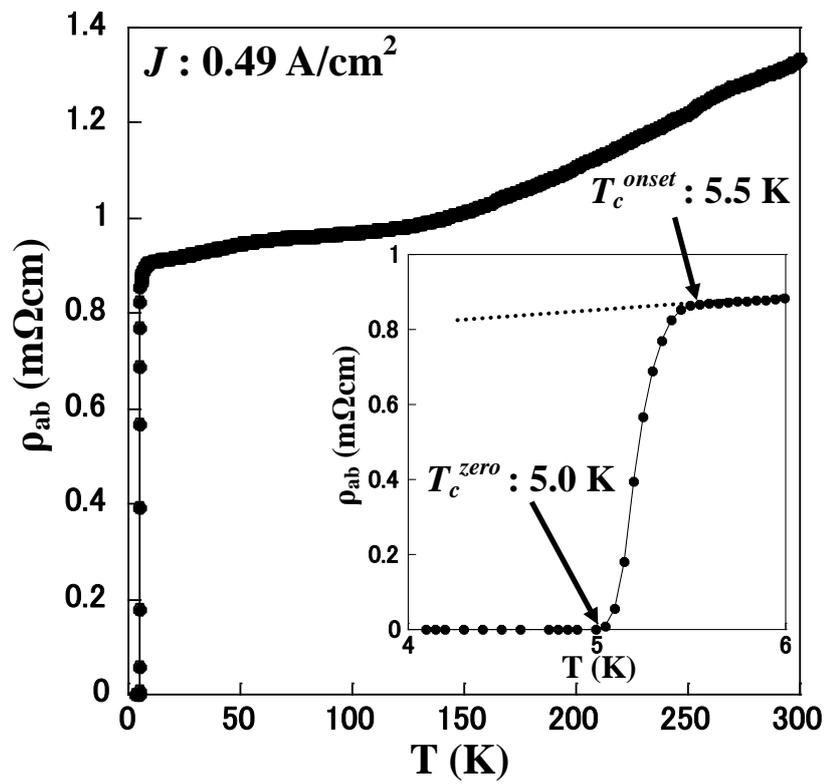

Figure 5

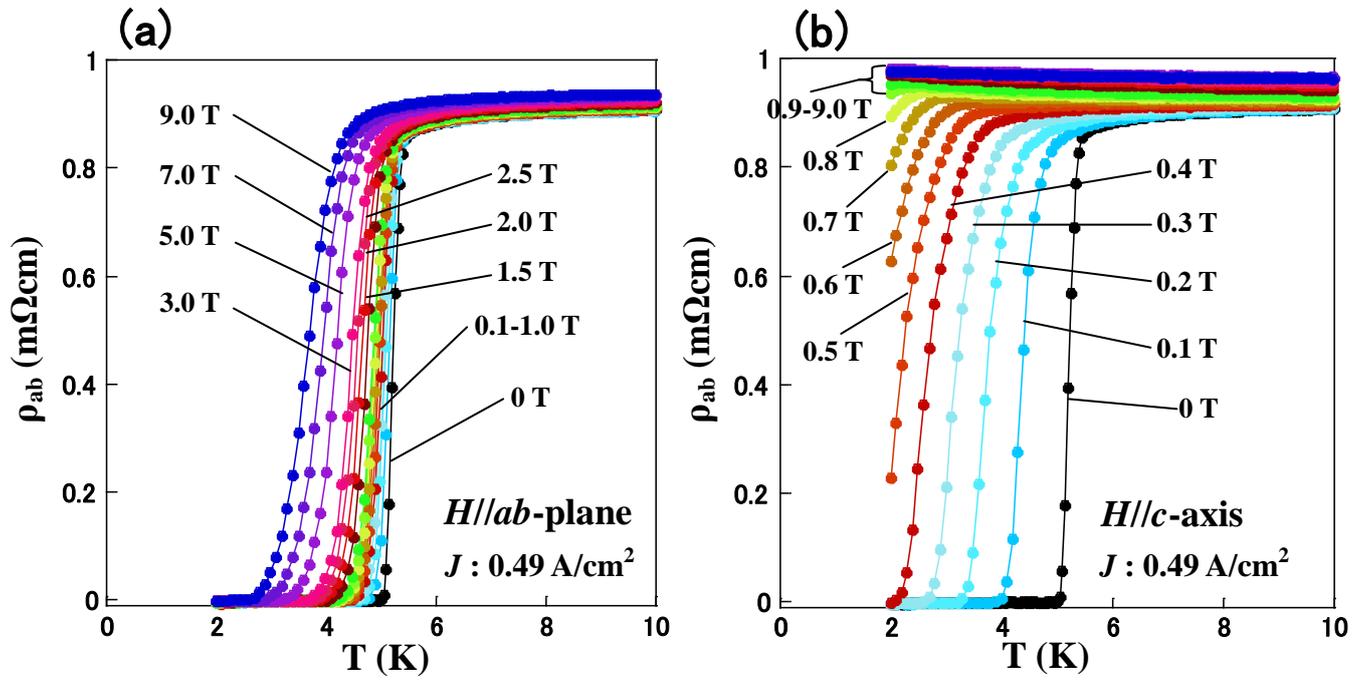

**Figure 6 (Color online)**

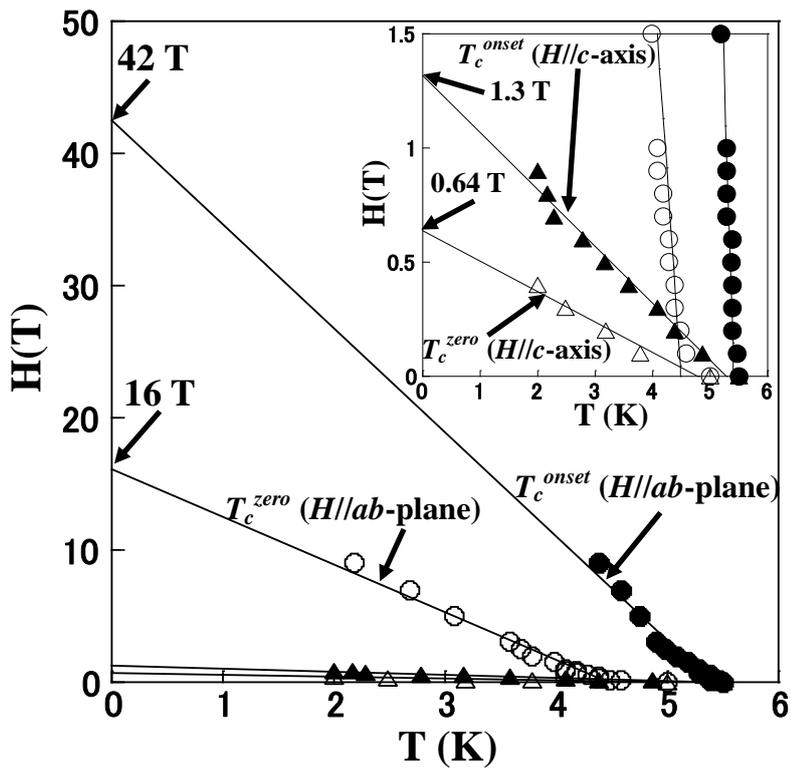

**Figure 7**

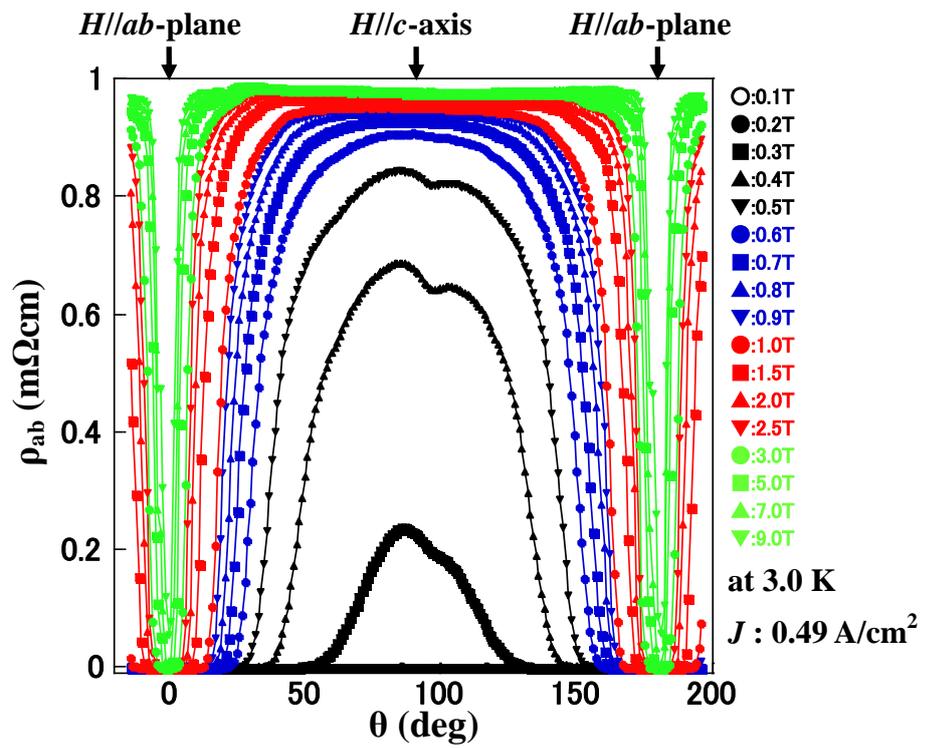

Figure 8 (Color)

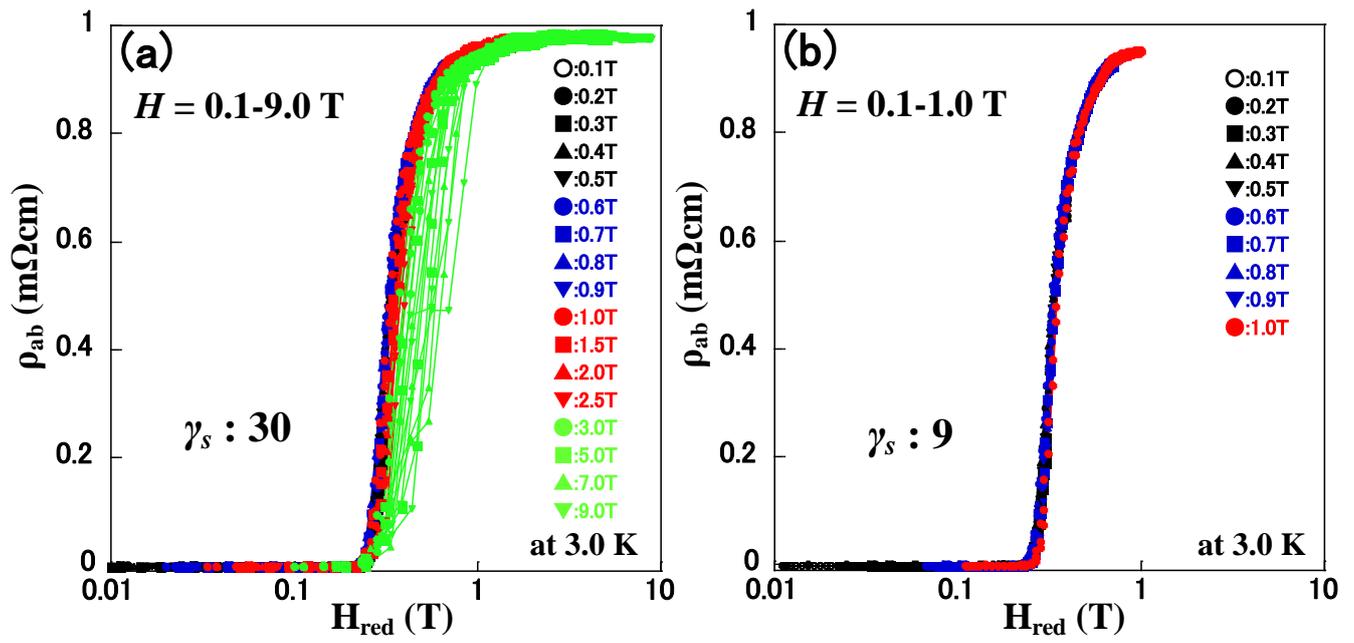

**Figure 9 (Color)**